\documentstyle[12pt]{article}
\begin{document}
\def\BR{I\hskip -4pt R}
\def\BC{I\hskip -7pt C}
\def\BT{I\hskip -5pt T}
\def\BZ{Z\hskip -6pt Z}
\def\Bz{Z\hskip -4pt Z}
\def\BN{I\hskip -4pt N}
\def\B1{1\hskip -4pt 1}
\baselineskip 22pt

\title
{On Group-Theoretic  Finite-Mode Approximation of  $2D$ Ideal 
Hydrodynamics
}
\author
{
Zbigniew Peradzy\'nski$^1$
 \and
Hanna E. Makaruk$^2$
\and Robert M. Owczarek$^2$\thanks{corresponding author: rmo@lanl.gov, 
Los Alamos NM 87545, USA}
}

\date{$^1$ University of Warsaw, Department of Mathematics, Banacha 2, 02-097 
Warsaw, Poland, and Institute of Fundamental Technological
Research, Polish Academy of Sciences, \'Swi\c{e}tokrzyska 21, 00-041
Warsaw, Poland
\\ $^2$   E-ET, 
MS M319, Los Alamos National Laboratory, Los Alamos,
NM 87545, USA,
on leave from Institute of Fundamental Technological Research,
Polish Academy of Sciences, \'Swi\c{e}tokrzyska 21,  00-041 Warsaw, Poland }
\maketitle

\begin{abstract}
      Structure constants 
      of the $su(N)$ ($N$ odd) Lie algebras 
      converge when N goes to infinity to the structure constants of
      the Lie algebra {\it sdiff}$(T^2)$ of the group of area-preserving
      diffeomorphisms of a $2D$ torus. Thus Zeitlin and others
      hypothesized that solutions of the Euler equations associated with
      $su(N)$ algebras converge to solutions of the Euler equations of 
      incompressible fluid dynamics on a $2D$ torus. In the paper we prove the hypothesis.
      Our numerical
      experiments show the Galerkin method applied to Euler equation 
	of hydrodynamics is computationally more efficient 
      in the range of time in which it is stable than that based on the SU(N)
	approximation.
	However, the latter is stable for 
	much longer time. These numerical results
      agree with theoretical expectations.

\end{abstract}

02.70.Dh --- Finite element and Galerkin methods\\
02.20.a --- Group theory\\
02.20.Tw --- Infinite dimensional Lie groups.

\section{Introduction}

       \paragraph{ }
      In the beginning of 90's a finite-mode
      approximation of $2D$ ideal hydrodynamics was introduced (\cite{zeitlin1}, \cite{dowker}, 
	\cite{abarbanel}).  The approximation is
      based on convergence of the structure constants of
      the Lie algebras $su(N)$ ($N$-odd), when $N \rightarrow \infty$
      to the structure constants of the Lie algebra
      {\it sdiff}$(T^2)$. The Lie algebra {\it sdiff}$(T^2)$ consists of divergenceless
      nonconstant
      vector fields on
      a $2D$ torus. The vector
      fields are infinitesimal generators of the group of area-preserving
      diffeomorphisms of the $2D$ torus. Incidentally, the group can be also identified as
      the group of symplectic diffeomorphisms, for the natural symplectic structure
      of the torus. The group is the basic symmetry group for incompressible ideal
      hydrodynamics on a $2D$ torus. In the physical model the fluid is maintained in a region of
      a plane, appropriate boundary conditions being satisfied. 
      The   
      approximation of {\it sdiff}$(T^2)$  by $su(N)$ is
      used quite widely both in hydrodynamics  and in string
      theory 
      \cite{zeitlin1,dowker,abarbanel,zeitlin3,hoppe,fairlie,floratos2,fairlie2}.
      There are also other finite-mode approximations of the Euler and similar equations 
      related to symplectic geometry 
      \cite{channel1,mclachlan,weinstein,ge}. 

		In \cite{schl} the approach to
      approximate {\it sdiff}$(T^2)$ is formalized and a weak uniqueness
      theorem is proved.  Unfortunately, the weak uniqueness does not lead to an
      isomorphism of the limit algebras.  Indeed, there are
      examples of various limit algebras which have their structure
      constants equal to the structure constants of some infinite
      dimensional algebras and despite this equality the limit algebras are not
      isomorphic to the infinite dimensional ones.
      As shown in \cite{schl}, the $\lim\limits_{N\to \infty} su(N)$ is neither
      isomorphic to {\it sdiff}$(T^2)$ nor to $sdiff(S^2)$, where
      $sdiff(S^2)$ consists of divergenceless vector fields on the 
      $2D$ sphere $S^2$. This approximation is still not well understood and additional work
      should be done towards its satisfactory formulation.

		The plan of the paper is as follows.

            In the second section 
      we discuss the issues connected with approximating the Lie algebra
      $sdiff(T^2)$ by the Lie algebras $su(N)$. We point out weaknesses 
      of this approximation. 
		
		In the third section we remind how the 
      Euler equations are being associated with arbitrary Lie algebras. Then we
      specify the Lie algebras to be $su(N)$ ($N$ odd) and $sdiff(T^2)$. 
	Despite the weak form of convergence of 
	$su(N)$ to {\it sdiff}$(T^2)$, solutions of the Euler 
	equations associated to the 
	Lie algebras $su(N)$ 
      converge, in a sense defined further in the paper, when $N$ goes to
      infinity, to the solutions of the Euler equation associated with the Lie algebra
      $sdiff(T^2)$. Proof of this fact is the central result of this paper.
      
		In the fourth section
      we discuss numerical calculations within this approach and within 
      the Galerkin approximation. Numerical stability 
      of this method in comparison to Galerkin one turns out to be significantly better.
      On the other hand, the numerical simulations demand significantly more calculation time
	than the Galerkin one.
      
		In the Conclusions we discuss the finite-mode
      approximation based on the group theoretic relations as a method of choice
	for solving Euler equation
      whenever numerical calculations are expected to give reliable results 
      for long evolution times, and adequate computational power
	is available.

		We supplement the paper with an Appendix, in which we discuss geometry of
	noncommutative tori, providing geometric interpretation of the approximation method.

\section{Approximation of the algebra {\it sdiff}$(T^2)$ by $su(N)$ algebras  }

      \paragraph{ }
      In this section we remind the formal approach to the
      approximation of {\it sdiff}$(T^2)$ and {\it sdiff}$(S^2)$ algebras given 
      in
      \cite{schl} and called there $L_\alpha$-convergence.  We remind here
      the result on weak uniqueness (proposition 3.3 from \cite{schl}).  We also
      discuss behavior of generators of $su(N)$ in the form chosen in
      \cite{zeitlin1} when $N$ goes to infinity.  We show that their norms
      diverge in this limit.

      In \cite{schl} instead of algebras $u(N)$, $su(N)$ are
      considered their complexifications, $gl(N, \BC)$ and $sl(N, \BC)$.  It is
      quite a standard to consider instead of algebras {\it sdiff}$(T^2)$,
      {\it sdiff}$(S^2)$, etc., their complexifications, so the above procedure
      is fully justified.  It does not lead to any principal differences
      in comparison with our dealing with corresponding real algebras,
      since the structure constants remain the same and the relations of
      being isomorphic or non-isomorphic are preserved by
      complexification.

      Let us begin with giving necessary definitions of objects appearing 
      in this section. First,
      let us remind some definitions and basic facts about the Lie algebras
      $gl(\infty)$, $gl_+(\infty)$, and $L_\Lambda$, as they are
      presented in \cite{schl}.

      First, $gl(\infty)$ is defined as
      $gl(\infty):=\{(a_{ij})_{ij\in Z\hspace{-1.5mm}Z}:a_{ij}\in
      C\hspace{-2.6mm}I$, all but finite number of $a_{ij}=0\}$.  This is
      then a matrix algebra with finite support.  As usual for matrix
      Lie algebras the commutator of matrices serves as the Lie bracket.

      The Lie algebra $gl_+(\infty)$ is defined in the same way with the
      only difference that the indices of the elements of the matrices
      in the Lie algebra are not integers $Z\hspace{-2mm}Z$
      but natural numbers $I\hspace{-1mm}N$.  Since
      $Z\hspace{-2mm}Z$ and $I\hspace{-1mm}N$ are bijective, any such
      bijection induces an isomorphism of $gl(\infty)$ and
      $gl_+(\infty)$.  However, the isomorphisms are not canonical, which
      means they depend on the choice of a basis in each of the Lie algebras.
      Since $gl(\infty)$ and
      $gl_+(\infty)$ have only finite number of elements different from
      zero, there is no problem with defining the trace of the elements
      in these Lie algebras. Namely, the number of non-zero 
      diagonal elements is also finite for all elements 
      of $gl(\infty)$ and $gl_+(\infty)$, and because of that the standard
      definition of a trace as the sum of diagonal elements of a matrix 
      is unchanged for both algebras.
      Then the Lie algebras $sl(\infty)$ and
      $sl_+({\infty})$ can be easily defined as consisting of those
      elements of $gl(\infty)$, $gl_+(\infty)$, which
      have trace equal to zero, and with the Lie structure inherited 
      from the algebras $gl(\infty)$, $gl_+(\infty)$, respectively.

      The Lie algebras {\it sdiff}$(T^2)$ and $L_\Lambda$, which are important
      in our further considerations,
      are defined in the
      following way.  One begins with the complex vector space:
\begin{equation}
V = \{T_{\bar m}, \bar m \in Z\hspace{-2mm}Z\times Z\hspace{-2mm}Z\},
\end{equation}
      it means a complex vector space generated by the basis $T_{\bar
      m}$, with indices $\bar m=(m_1,m_2)$ such that $m_1$ and $m_2$ are
      integers.

      The following interesting Lie algebra structures, 
      indexed by $\Lambda\in I\hspace{-1mm}R$, can be defined on $V$:  
      For $\Lambda \in
      I\hspace{-1mm}R-\{0\}$ they are defined on the generators $T_{\bar m}$
      by:
\begin{equation}
[T_{\bar m},T_{\bar n}]^\Lambda :=
\frac{1}{2\pi\Lambda}\sin
{\left(2\pi\Lambda(\bar m\times \bar n)\right)}T_{\bar m + \bar n}
\end{equation}
      and for $\Lambda=0$ by:
\begin{equation}
[T_{\bar m},T_{\bar n}]^o := (\bar m \times \bar n)T_{\bar m + \bar n},
\end{equation}
      where
\begin{equation}
\bar m\times \bar n = m_1 n_2 - m_2 n_1
\hspace{1cm}\mbox{for}\hspace{1cm}
\bar m = (m_1,m_2),\,\,\,\,\bar n = (n_1,n_2).
\end{equation}
      Then, as usual, the Lie brackets are extended to the whole $V$   by linearity.
      These Lie algebras are denoted $\widetilde
      L_\Lambda=(V,[\cdot,\cdot]^\Lambda)$.  They are direct sums of
      subalgebras
\begin{equation}
\widetilde L_\Lambda = \{T_{(0,0)}\} \oplus
\left\{T_{\bar m}:\bar m \in Z \hspace{-2mm}
Z \times Z \hspace{-2mm} Z - \{0\}\right\}
\end{equation}
      where $\{\cdot \}$ denotes a vector space spanned on generators listed
      inside the figure brackets, and the Lie structure is defined by 
      restriction of $[\cdot,\cdot]^\Lambda$ to the given subspace of $V$.

      The Lie algebra $L_\Lambda$ is defined as the second summand in
      the above sum, 
      $L_\Lambda = (\left\{T_{\bar m}:\bar m \in Z \hspace{-2mm}Z
      \times Z \hspace{-2mm} Z - \{0\}\right\}), [\cdot,\cdot]^\Lambda)$.
      In the case $\Lambda=0$, $L_o$ is identified with
      the complexified Lie algebra of the nonconstant divergenceless
      vector fields on $T^2$, which is the {\it sdiff}$(T^2)$.

      Since approximation of the Lie algebras $L_\Lambda$, {\it sdiff}, etc.
      by the algebras $gl(n)$ or $su(n)$ can not be realized in a strong
      sense of inductive or projective limit as it is often done for
      other infinite-dimensional algebras, the authors of \cite{schl}
      axiomatized the concept of so called $L_\alpha$-limit of the Lie
      algebras.  This notion is a generalization of intuitively defined
      limits from the papers \cite{fairlie2}, \cite{pope}.

      The definition consists of two axioms.  The first axiom
      demands existence of a special family of real or complex Lie algebras
      $({\cal G}_\alpha, [\cdot,\cdot]_\alpha, \alpha \in I)$, where the
      index set $I=I\hspace{-1mm}N$ or $I\hspace{-1mm}R$.  This family
      should be equipped with a family of metrics $d_\alpha$.  Accordingly to
      the axiom there should exist another real or complex Lie algebra
      $({\cal G},[\cdot,\cdot])$ such that for every $\alpha \in I$
      there exists a surjective map $p_\alpha:  \,\,{\cal G}\mapsto
      {\cal G}_\alpha$ and the condition
\begin{equation}
\forall x,y \in {\cal G} \mbox{ if }
d_\alpha(p_\alpha(x),p_\alpha(y)) \longrightarrow_{\hspace{-7mm}
\alpha\to\infty}0,\mbox{ then } x = y.
\end{equation}
      should be satisfied.
      The second axiom concerns the  definition of a Lie algebra structure on the
      $L_\alpha$-limit.  Accordingly to this axiom $({\cal G},
      [\cdot,\cdot])$, is called $L_\alpha$-limit of the algebras
      $({\cal G}_\alpha, [\cdot,\cdot]_\alpha,d_\alpha, \alpha \in I)$
      and the family of the algebras is called the approximating
      sequence introduced by $(p_\alpha, \alpha\in I)$ if the condition
\begin{equation}
d_\alpha(p_\alpha[x,y],[p_\alpha x,p_\alpha y]_\alpha)
\longrightarrow_{\hspace{-7mm}\alpha\to\infty}0.
\end{equation}
      is satisfied.
      There (in \cite{schl}) was also proved a weak uniqueness theorem, which states that
      if $({\cal G}_\alpha, [\cdot,\cdot]_\alpha,d_\alpha, \alpha \in
      I)$ is an approximating sequence for $({\cal G},[\cdot,\cdot])$
      induced by $(p_\alpha, \alpha\in I)$, and ${\cal G}'$ is a linear
      subspace of ${\cal G}$, which carries a Lie structure $[\cdot,
      \cdot]'$ which projects on each ${\cal G}_\alpha$, then $({\cal
      G}',[\cdot,\cdot]')$ is a Lie subalgebra of $({\cal
      G},[\cdot,\cdot])$ if and only if the approximating sequence for
      ${\cal G}$ is by restriction also approximating sequence for
      ${\cal G}'$ induced by the restriction of the $p_\alpha$.

      However, this theorem about weak uniqueness is not of very big value, 
      because it does not guarantee
      that if we start with a given approximating sequence of the Lie
      algebras $({\cal G}_\alpha, \alpha \in I)$ we end up with
      isomorphic limit algebras, independently of how we came to the limits.

      Let us remind how {\it sdiff}$(T^2)$ is described rigorously within the
      $L_\alpha$ limit of the algebras $L_\Lambda$ when $\Lambda\to 0$.
      One begins with the sequence of algebras $L^N=\widetilde
      L_{1/N}$, i.e. the sequence of algebras $\widetilde L_{\Lambda}$
      with $\Lambda=\frac{1}{N}$.  One distinguishes an ideal $J^N$ in
      each algebra $L^N$ of the sequence:
\begin{equation}
J^N := \{T_{\bar m} - T_{\bar m + N\bar a}:\,\,\,
\bar m, \bar a \in  Z \hspace{-2mm} Z \times  Z \hspace{-2mm} Z \}.
\end{equation}
      Then one defines the sequence of factor algebras
      $L_{(N)}:=L^N/J^N$ with $\pi_N:L^N\to L_{(N)}$ the canonical
      projection.  As the basis of each algebra in the sequence serve
      $\pi_N(T_{\bar m}),\,\,\,\,\bar m=(m_1,m_2),\,\,\,\,0\leq
      m_1,m_2<N$.  The Lie structure in each algebra of the sequence is
      given by:
\begin{equation}
[\pi_N(T_{\bar m}),\pi_N(T_{\bar n})]^{(N)} =
\frac{N}{2\pi}\sin{\left(\frac{2\pi}{N}
(\bar m \times \bar n)\right)}\,\pi_N(T_{(\bar m + \bar n) mod\, N}).
\end{equation}

      This algebra $L_{(N)}$ is exactly identical with the algebra $gl(N)$ with $N$
      odd.  The approximating sequence consists then of the algebras
      $(L_{(N)}, [.,.]^{(N)})$ with the index set
      $I=I\hspace{-1mm}N$. The surjective maps $p_\alpha$ are 
      identified in this case as the maps $\pi_N$.  The
      metric on each $L_{(N)}$ is induced by defining it on the generators by
\begin{equation}
<\pi_N(T_{\bar m}),\pi_N(T_{\bar n})> = \delta_{m_1,n_1}\delta_{m_2,n_2}.
\end{equation}
      and then extending to the whole $L_{(N)}$ by linearity.

      It is shown in \cite{schl} that the 2 axioms of $L_\alpha$-limit
      are satisfied for this sequence.  Therefore, the algebra {\it sdiff}$
      (T^2)$ is an $L_\alpha$-limit of the algebras $gl(N, \BC)$.
      Nevertheless, $gl(\infty)$ which is an inductive limit of the
      algebras $gl(N, \BC)$, so it is the limit in a strong sense, is not
      isomorphic to {\it sdiff}$(T^2)$.  This fact is also proved in 
      \cite{schl}.  

      The above example shows that one should be careful with approximating
      infinite-dimensional real algebras by finite dimensional ones in
      such weak sense as the $L_\alpha$-limit. 
      We will demonstrate now another difficulty of this definition
      of a limit. The difficulty comes from the fact 
      that the renormalized basis of the Lie algebra $su(N)$ 
      has divergent norm for $N$ going to infinity. To be more precise, 
      let us remind that 
      the
      basis of the $su(N)$ algebra which leads to the mentioned
      $L_\alpha$ -approximation of {\it sdiff}$(T^2)$ is defined as:
\begin{equation}
{\cal J}_{\bar n} = \lambda^{n_1n_2/2}g^{n_1}h^{n_2},
\label{19ow}
\end{equation}
where $\lambda=e^{4i\pi/N}$,
$g=diag(1,\lambda,\ldots,\lambda^{N-1})$, and 

$$h=\left[\begin{array}{cccccc}0&1&.&.&.&0 \\
0&0&1&.&.&0\\
.&.&.&.&.&.\\
1&.&.&.&.&0 \end{array}\right]$$
      is a cyclic permutation matrix.  Then, to obtain desired form of
      the structure constants of the algebra, a renormalization is done
      by multiplying all the basis elements by $\frac{iN}{4\pi}$.  As a
      result, the renormalized basis elements for $N\to\infty$ become
      elements of infinite norm.  This is in apparent contrast to the
      behavior of the basis elements of the algebra {\it sdiff}$(T^2)$
\begin{equation}
L_{\bar n} = i \exp{(i\bar n\cdot \bar x)}(n_2\partial_1 - n_1\partial_2)
\label{20ow}
\end{equation}
      , which are all of finite norm.

      It seems then that the algebraic arguments on approximation of
      {\it sdiff}$(T^2)$ by $su(N)$ need careful
      treatment.

      What is also of interest, especially in the context of geometric description of 
      an ideal incompressible fluid, and its geometric quantization  like in \cite{GMS1} 
      or \cite{GOS}, it is a better understanding of convergence of coadjoint 
      orbits of the groups $SU(N)$ to the coadjoint orbits of the goup 
      $SDiff(T^2)$. Analogous interesting question  
      concerns the Casimir functions. These in the case of $SU(N)$ are: 
      $tr\,A^2$,  $tr\,A^3$, ...  $tr\,A^n$,... for an $A\in su(N)$. 
      Their counterparts for the $SDiff(T^2)$ are: 
      $\int_{T^2}\omega^2\mu$, 
      $\int_{T^2}\omega^3\mu$,...,$\int_{T^2}\omega^n\mu,...,$ 
      where $\mu$ is the standard surface element of the torus $T^2$.
      The work on this issue is in progress.

\section{Approximating  Euler equation}

\paragraph{ }
      In this section we introduce the Euler equation for incompressible
      flows of an ideal fluid on a $2D$ torus. Then we remind the definition 
      of the Euler equations connected with arbitrary Lie algebras, due to Arnold.
      Next, we specify the latter equations to the case of $su(N)$ Lie algebras
      ($N$ odd) and to the case of the Lie algebra $sdiff(T^2)$. Finally, we prove 
      the theorem on convergence of solutions
      of the Euler equations for $su(N)$ to the solutions of the Euler equation
      connected with the $sdiff(T^2)$ Lie algebra.

      As was noticed in \cite{arnold}, the Euler equations can be
      generalized to any Lie algebra by
\begin{eqnarray}
{\dot {\omega }}_{i} = g^{jk} C^{r}_{ik} \omega _{j} \omega _{r},
\label{4a}
\end{eqnarray}
      where $C^{r}_{ik} $ are structure constants of the algebra and
      $g$ is a symmetric metric. For example, g is the inverse  of the inertia tensor in the
      case of motion of a rigid body, for which the appropriate algebra is 
      $so(3)$). The summation over repeating indices is understood.

      The equation for two-dimensional hydrodynamics of an ideal fluid
      can be expressed in terms of the Poisson bracket $\{\cdot,\cdot\}$ of 
      vorticity
      $\omega $ and the stream function as follows
\begin{eqnarray}
\dot \omega  = \left\{\Delta ^{- 1}\omega ,\omega \right\},
\label{1a}
\end{eqnarray}
      where $\Delta ^{- 1}\omega $ denotes the stream function, i.e. the
      solution of $\Delta \psi = \omega $.
      In the case of the flow in a
      rectangle with periodic boundary condition (i.e. on a two-dimensional
      torus $T^{2}$) we assume the mean vorticity equal to zero,
      $\int _{T^{2}} \omega = 0$.  As noticed by Zeitlin, the symplectic
      diffeomorphisms of $T^{2}$ are generated by
\begin{eqnarray*}
L_{\bar n} = i \exp [i(n_{1}x_{1} + n_{2}x_{2} )]
(n_{2}\partial _{1} - n_{1}\partial _{2}) =
-i \exp (i \bar n \cdot \bar x) (\bar n \times \nabla ),
\end{eqnarray*}
      where $\bar n^{2} = (n^{2}_{1} + n^{2}_{2}) \neq 0$.  These operators
      satisfy the Jacobi identity and form the Lie algebra $sdiff(T^{2})$.
      For the commutators we have
\begin{eqnarray}
[L_{\bar n},L_{\bar m}] = (\bar n \times \bar m) L_{\bar n+\bar m}, \qquad
\hbox {where~~~~} \bar n \times \bar m = n_{1}m_{2} - n_{2}m_{1}.
\label{2a}
\end{eqnarray}
      Equation (\ref{1a}), when expressed in the basis (\ref{20ow}),
      i.e. in terms of the Fourier components, $\left\{\omega_{\bar n}\right\}$ 
      of $\omega$, takes the form
\begin{eqnarray}
{\dot \omega} _{\bar m} =
\sum _{\bar k} \frac {\bar m \times \bar k}{\bar k^{2}} \omega _{\bar m+\bar k} 
\omega _{-\bar  k},
\label{3a}
\end{eqnarray}
      where $\omega _{- \bar k} = \omega _{\bar k}^{*}$, because $\omega(\bar  
      x)$ is real.  
      Taking (\ref{19ow}) as the basis
      for the Lie algebra $su(N)$ for $N = 2n + 1$ one can construct
      a system which formally approximates the system (\ref{3a}).
      Assuming $L_{\bar n} = \frac {iN}{4\pi } {\cal J}_{\bar n}$, where
      ${\cal J}_{\bar n}$ are defined in (\ref{19ow}) we have
\begin{eqnarray}
[L_{\bar n},L_{\bar m}] = \frac {N}{2\pi}
\sin [(\frac {2\pi}{N})(\bar n \times\bar  m)] L_{_{\bar n+\bar m\,|{\rm mod}N}}
\label{5a}
\end{eqnarray}
      in which the summation $modN$ in both components is understood. 
      Thus, assuming the inertia
      tensor analogous to the one in (\ref{3a}) we may define the following
      Euler equations on the dual of $su(N)$
\begin{eqnarray}
{\dot \omega}_{\bar m} = \sum _{\bar k}
\frac {\sin [2\pi (\bar m \times\bar  k)/N]}{\frac {2\pi }{N}\bar k^{2}}
\omega _{_{\bar m+\bar k\,|{\rm mod}N}} \omega _{- \bar k},
\label{6a}
\end{eqnarray}
      where $- \frac {N-1}{2} \leq k_{i}, m_{i}\leq \frac {N-1}{2}$
      , $i = 1,2$. In the literature the system (\ref{6a}) is
      known as a finite dimensional analog of two-dimensional
      hydrodynamics. However, no proof was known that solutions of
      (\ref{6a}) are indeed approximating those of (\ref{3a}). To prove
      that such a correspondence exists we will treat solutions of
      (\ref{6a}) as infinite sequences assuming $\omega _{\bar k}$,
      $\omega _{-\bar  k}$ are zero for $\mid k_{i}\mid > \frac {N-1}{2}$.
      Let us assume therefore that $\left\{\omega _{\bar k}(t)\right\}$ is a
      solution of Eqs(\ref{3a}) with the initial data
      $\left\{\omega _{\bar k}(0)\right\}$. Correspondingly, let
      $\left\{\tilde {\omega} _{\bar k}(t)\right\}$ be a solution of
      Eqs(\ref{6a}) with initial values
      $\left\{{\tilde \omega }_{\bar k}(0)\right\}$.  We will assume that
      the difference $\omega _{\bar k}(0) - \tilde {\omega }_{\bar k}(0)$ is 
      small
      in $l^{2}$ i.e.  $\sum _{\bar k} \mid \omega _{\bar k}(0) -
      \tilde {\omega }_{\bar k}(0) \mid ^{2} < \delta $.
      Then we will prove under some additional assumptions  for
      $\omega $ that on a finite time interval the solutions are also
      close. Let us write the equation for the difference
      $\Omega(t) = \omega (t) - {\tilde \omega }(t)$. Subtracting (\ref{6a})
      from (\ref{3a}) one obtains after some manipulations
\begin{eqnarray}
{\dot{\Omega }}_{\bar m} = \sum _{\bar k} \frac {\bar m \times \bar k}{\bar 
k^{2}}
\left(1 - \frac {\sin (\varepsilon\bar  m \times \bar k)}{\varepsilon \bar m 
\times \bar k}
\right) \omega _{\bar m+\bar k} \omega _{- \bar k} +\nonumber \\
\sum _{\bar k} \frac {\sin (\varepsilon \bar m \times \bar k)}{\varepsilon \bar 
k^{2}}
\left\{\Omega _{\bar m+\bar k} \omega _{- \bar k} + {\tilde \omega }_{\bar 
m+\bar k} \Omega _{- \bar k}
\right\},
\label{7a}
\end{eqnarray}
      where $\varepsilon = \frac {2\pi}{N}$. It is desirable to express the 
r.h.s. of (\ref{7a})  
     only in terms of variables $\Omega_{\bar k}$, $\omega_{\bar k}$. Since     
$\omega _{\bar m+\bar k} =$ $ {\tilde \omega }_{\bar m+\bar k} + \Omega _{\bar 
m+\bar k}$, we have
\begin{eqnarray}
{\dot{\Omega }}_{\bar m} = \sum _{\bar k} \frac {\bar m \times \bar k}{\bar 
k^{2}}
\left(1 - \frac {\sin (\varepsilon\bar  m \times \bar k)}{\varepsilon \bar m 
\times \bar k}
\right) \omega _{\bar m+\bar k} \omega _{- \bar k} +\nonumber \\
\sum _{\bar k} \frac {\sin (\varepsilon \bar m \times \bar k)}{\varepsilon \bar 
k^{2}}
\left\{\Omega _{\bar m+\bar k} \omega _{- \bar k} + {\omega }_{\bar m+\bar k} 
\Omega _{- \bar k}-
\Omega_{\bar m+\bar k} \Omega _{- \bar k}
\right\},
\label{*}
\end{eqnarray}
%

%
%

Now we are ready to derive the equation for the $L^{2}$-norm of $\Omega $
\begin{eqnarray}
\frac {1}{2} \parallel \Omega \parallel  ^{2} =\frac {1}{2} \sum _{\bar m} \mid 
\Omega _{\bar m}\mid ^{2}=
\frac {1}{2} \sum _{\bar m} \Omega _{\bar m} \Omega _{-\bar  m}.
\label{9a}
\end{eqnarray}
The last equality in the  chains follows from $\Omega(\bar x)$ being real.

Let us observe that 
\begin{equation}
\frac{1}{2}\frac{d}{dt}\sum _{\bar m} \Omega _{\bar m} \Omega _{-\bar  
m}=\frac{1}{2}
\sum_{\bar m}\left(\frac{d\Omega_{\bar m}}{dt}\Omega_{-\bar m}+\Omega_{\bar 
m}\frac{d\Omega_{-\bar m}}{dt}\right)
=\sum_{\bar m}\Omega_{\bar m}\frac{d\Omega_{-\bar m}}{dt}
\end{equation}
,since the sum $\sum_{\bar m}\Omega_{\bar m}\frac{d\Omega_{-\bar m}}{dt}$ by 
substitution $\bar m\to -\bar m$ becomes 
$\sum_{\bar m}\Omega_{-\bar m}\frac{d\Omega_{\bar m}}{dt}$.

Multiplying Eq.(\ref{*}) by $\Omega _{-\bar  m}$, performing the summation
over $\bar m$  one obtains
\begin{eqnarray}
\frac {1}{2} \frac {d}{dt}\parallel  \Omega (t)\parallel  ^{2} &=&
\sum _{\bar m} a_{\bar m}(\varepsilon ) \Omega _{-\bar m} \nonumber\\ &+&
\sum _{\bar m} \sum _{\bar k} \frac {\sin (\varepsilon \bar m \times \bar 
k)}{\varepsilon\bar  k^{2}}
\omega _{\bar m+\bar k} \Omega _{- \bar k} \Omega _{-\bar  m}\nonumber\\ &+&
\sum _{\bar m} \sum _{\bar k} \frac {\sin (\varepsilon \bar m \times \bar 
k)}{\varepsilon\bar  k^{2}}
\Omega _{\bar m+\bar k} \Omega _{- \bar m} \omega _{-\bar  k}\nonumber\\ 
&-&\sum _{\bar m} \sum _{\bar k} \frac {\sin (\varepsilon \bar m \times \bar 
k)}{\varepsilon\bar  k^{2}}
\Omega _{\bar m+\bar k} \Omega _{- \bar m} \Omega _{-\bar  k}
\label{10a}
\end{eqnarray}
where
\begin{eqnarray}
a_{\bar m}(\varepsilon ) = \sum _{\bar k} a^{\bar k}_{\bar m} (\varepsilon )
\omega _{\bar m+\bar k} \omega _{- \bar k}.
\label{11a}
\end{eqnarray}
and where
\begin{eqnarray}
a^{\bar k}_{\bar m}(\varepsilon ) = \frac {\bar m \times \bar k}{\bar k^{2}}
\left(1 - \frac {\sin (\varepsilon \bar m \times \bar k)}{\varepsilon \bar m 
\times\bar  k}
\right).
\label{12a}
\end{eqnarray}
In the last two sumands repeats the term:

\begin{equation}
I_{\bar k}:=\sum _{\bar m} \sin (\varepsilon \bar m \times \bar k)\Omega _{\bar 
m+\bar k} \Omega _{- \bar m} 
\label{12b}
\end{equation}

Changing the name of the summation index does not change the  value of $I_{\bar 
k}$. Let us then introduce the change from $\bar m$ into $-(\bar m +\bar k)$. As 
a result one gets ($\bar m '=-(\bar m +\bar k)$ and then we omit $'$).
\begin{eqnarray}
I_{\bar k}&=&\sum _{\bar m} 
\sin (\varepsilon (-(\bar m +\bar k)\times \bar k))\Omega _{- \bar m}\Omega 
_{\bar m+\bar k}\nonumber\\
 &=&  -\sum _{\bar m}\sin (\varepsilon (\bar m \times \bar k)\Omega _{\bar 
m+\bar k}\Omega _{- \bar m}=-I_{\bar k} 
\label{12c}
\end{eqnarray}
Therefore $I_{\bar k}\equiv 0$.

Now we will prove that the sequence
$\left\{a_{\bar m}(\varepsilon )\right\}$ defined by (\ref{11a})
satisfies $\sum \mid a_{\bar m} \mid ^{2} = \delta (\varepsilon ) \rightarrow 0 
$
~when $~\varepsilon \rightarrow 0~$ provided that the solution $\omega (x)$
belongs to the Sobolev space\footnote{We know \cite{lions} that in
two-dimensional case the Euler equations of an ideal fluid have global
solutions whose quality (differentiability properties) are related
to the quality of initial data.}
 $H^{\sigma } (T^{2})$ for $\sigma > 2$.
We have in such a case
$\sum _{\bar m} \mid \omega _{\bar m}\mid ^{2} (1 + \bar m^{2})^{\sigma} < 
\infty $
and therefore $\omega _{\bar m}$ can be represented as
$\omega _{\bar m} = \frac {c_{\bar m}}{(1 + \bar m^{2})^{\sigma/2}}$ where still
$\sum _{\bar k} \mid c_{\bar k}\mid ^{2} = c < \infty $. Thus denoting
$\beta ^{\bar k}_{\bar m}(\varepsilon ) =a^{\bar k}_{\bar m}(\varepsilon ) 
(1 + (\bar m + \bar k)^{2})^{- \sigma/2 } (1 + \bar k^{2})^{- \sigma/2 }$
we have $a_{\bar m} =$ $\sum _{\bar k} \beta ^{\bar k}_{\bar m}(\varepsilon ) 
c_{\bar m +\bar  k} c_{-\bar k}$
,which by Schwartz inequality gives us
\begin{eqnarray}
\mid a_{\bar m}\mid \leq
\sqrt {\sum _{\bar k} \mid \beta ^{\bar k}_{\bar m} (\varepsilon ) \mid ^{2} }~
\sqrt {\sum _{\bar k} \mid c_{\bar m+\bar k} c_{-\bar  k} \mid ^{2} }.
\label{13a}
\end{eqnarray}
The last sum on the right hand side is obviously bounded because
\begin{eqnarray*}
\sum _{\bar k} \mid c_{\bar m+\bar k} c_{- \bar k}\mid ^{2} \leq
{\rm sup}_{_{_{_{\hskip -10pt{\bar m}}}}} \mid c_{\bar m}\mid ^{2}
\sum _{\bar k} \mid c_{\bar k}\mid ^{2} = C_{1} < \infty .
\end{eqnarray*}
Therefore as it follows from (\ref{13a})
\begin{eqnarray}
\sum _{\bar m} \mid a_{\bar m} \mid ^{2} \leq
C_{1}\,\sum _{\bar m} \sum _{\bar k} \mid \beta ^{\bar k}_{\bar m} (\varepsilon 
)\mid ^{2}.
\label{14a}
\end{eqnarray}
\par To prove that this double sum tends to zero for
$\varepsilon \rightarrow 0$, we use the integral criterion; we denote
\begin{eqnarray*}
G(\varepsilon ) = \int _{\mid \bar k\mid > 1}dk_{1}\, dk_{2}
\int _{\mid \bar m+\bar k\mid > 1}dm_{1}\, dm_{2} \mid \beta ^{\bar k}_{\bar m} 
(\varepsilon )\mid ^{2}
.
\end{eqnarray*}
By introducing new variables $s_{1} = (\bar m + \bar k) \times \frac {\bar 
k}{\mid\bar  k\mid }$;
$s_{2} = (\bar m + \bar k) \times \frac {\bar k^\bot}{\mid\bar  k\mid }$, where
$\bar k^\bot = (- k_{2}, k_{1})$ we estimate $G(\varepsilon )$ extending
integration in $s_{1}$, $s_{2}$ on the whole $\BR^{2}$
\begin{eqnarray*}
G(\varepsilon ) \leq \int _{I\hspace{-.2mm}R^{2} \setminus D} dk_{1}\,dk_{2}
\int _{I\hspace{-.2mm}R^{2}} \left( 1 -
\frac {\sin (\varepsilon \mid\bar  k\mid s_{1})}{\varepsilon \mid \bar k\mid 
s_{1}}
\right)^{2} \frac {s^{2}_{1}}{\bar k^{2}}
\frac {ds_{1}\, ds_{2}}
{(1 + s_{1}^{2} + s_{2}^{2})^{\sigma } (1 + \bar k^{2})^{\sigma }},
\end{eqnarray*}
where $D$ is the unit disc on $\BR^{2}$, $\bar k^{2} < 1$. For $\sigma > 2$
the integral over $s_{2}$ is finite and we have
\begin{eqnarray*}
\int _{I\hspace{-.2mm}R} \frac {ds_{2}}{(1 + s_{1}^{2} + s_{2}^{2})^{^{\sigma}}} 
=
\frac {M _{\sigma }}{(1 + s^2_{1})^{^{\sigma - 1/2}}},
\end{eqnarray*}
where the constant $M_{\sigma }$ grows to infinity for
$\sigma \rightarrow 2$.
Therefore introducing $z = \varepsilon \mid \bar k\mid s_{1}$ we have
\begin{eqnarray*}
G(\varepsilon )\leq \frac {M_{\sigma }}{\varepsilon ^{3}}
\int _{I\hspace{-.2mm}R^{2}\setminus D} \frac {dk_{1}\, dk_{2}}{\bar k^{2}(1 + 
\bar k^{2})^{\sigma }}
\int _{I\hspace{-.2mm}R^{1}} (1 - \frac {\sin z}{z})^{2}
\frac {z^{2} \, dz}{(1 + \frac {z^{2}}{\varepsilon ^{2}\bar k^{2}})^{\sigma 
-1/2}}.
\end{eqnarray*}
Noting that
$\mid 1 - \frac {\sin z}{z}\mid \leq {\rm min~}(\frac {z^{2}}{6}, 2)$
we can estimate the integral over $dz$ splitting it in two parts
as follows
\begin{eqnarray*}
\int _{0}^{\infty } (1 - \frac {\sin z}{z})^{2}
\frac {z^{2} \, dz}{(1 + \frac {z^{2}}{\varepsilon ^{2}\bar k^{2}})^{^{\sigma 
-1/2}}}
\leq \frac {1}{36 }
\int _{0}^{\mid \bar k\varepsilon \mid ^{\alpha}} z^{6} \,dz+
\left\vert 4\int ^{\infty }_{\mid\bar  k\varepsilon \mid ^{\alpha}}
\mid \varepsilon\bar  k\mid ^{2\sigma - 1} z^{3-2\sigma } dz\right\vert =\\
=\frac {1}{252} \mid \bar k\varepsilon \mid ^{7\alpha } +
\frac {2}{\mid 2 - \sigma \mid}
\mid \varepsilon \bar k\mid ^{2\sigma -1+2\alpha (2-\sigma )}.
\end{eqnarray*}
Taking $\alpha = \frac {2\sigma - 1}{2\sigma + 3}$ we make both exponents
equal to $7\, \frac {2\sigma - 1}{2\sigma + 3}$ and finally one obtains
\begin{eqnarray*}
G(\varepsilon ) \leq \varepsilon ^{r} M_{\sigma }
\left\{\frac {1}{126} + \frac {4}{ \sigma -2}\right\}
\int _{I\hspace{-.2mm}R^{2}\setminus D}
\frac {\mid\bar  k\mid ^{7\alpha -4}}{(1 + \bar k^{2})^{\sigma }} dk^{1}\, 
dk^{2},
\end{eqnarray*}
where $r = 8 \frac {\sigma - 2}{2\sigma + 3}$. For $\sigma > 2$,
the integral over $k^{1}$, $k^{2}$ is finite, and thus
$G(\varepsilon ) \rightarrow 0$ for $\varepsilon \rightarrow 0$.

In this way we have proved that for $\sigma > 2$ the first term
on the r.h.s. of Eq(\ref{10a}) is vanishing when $N \rightarrow \infty $
($\varepsilon = 2\pi /N$). It converges to zero as $\varepsilon ^{r/2}$,
i.e. as $N^{- r/2}$, where $r$ depends on $\sigma $. Indeed, we have
\begin{eqnarray*}
\mid \sum _{\bar m} a_{\bar m}(\varepsilon )\Omega_{-\bar m} \mid \leq
\mid \Omega \mid \left(\sum _{\bar m} \mid a_{\bar m}(\varepsilon )\mid ^{2}
\right)^{1/2} = \mid \Omega \mid G(\varepsilon )^{1/2}.
\end{eqnarray*}
     Now let us estimate the second term on the r.h.s of (\ref{10a}).
     Denoting $\bar \mu =\bar  m + \bar k$ we have
\begin{eqnarray*}
\mid \sum _{\bar m} \sum _{\bar k}
~\frac {\sin (\varepsilon \bar m \times\bar  k)}{\varepsilon\bar  k^{2}}~
\omega _{\bar m+\bar k}\,\Omega _{- \bar k}\,\Omega _{- \bar m} \mid \leq 
\sum _{\bar \mu,\bar k} \mid
~\frac {\sin (\varepsilon \bar \mu \times \bar k)}{\varepsilon (\bar \mu 
\times\bar  k)}~
\frac {\bar \mu \times\bar  k}{\bar k^{2}}~
\frac {c_{\bar \mu}}{(1 +\bar  \mu^{2})^{\sigma /2}}~
\Omega _{-\bar  k}\,\Omega _{\bar k - \bar \mu} \mid \leq \\
\sum _{\bar \mu ,\bar k} \,\mid c_{\bar \mu}\,(1 + \bar \mu^{2})^{^{ {1-\sigma 
}\over {2} }}\mid
~\mid \Omega _{-\bar  k}\,\Omega _{\bar k - \bar \mu}\mid \leq
K_{\sigma }\,\parallel \Omega \parallel ^{2},
\end{eqnarray*}
     where $K_{\sigma } =
     \sum_{\bar \mu}\,\mid c_{\bar \mu}\,(1 + \bar \mu^{2})^{^{ {1-\sigma }\over {2} 
     }}\mid$.
      From Schwartz inequality $K_{\sigma}$ is finite for $\sigma > 2$.

      Now let us come back to Eq.(\ref{10a}). Dividing both sides by
      $\parallel \Omega (t)\parallel$ and integrating with respect to time we come to
\begin{eqnarray*}
\parallel \Omega (t)\parallel = \parallel \Omega (0)\parallel +
\int _{0}^{t} \frac {dt }{\parallel \Omega (t)\parallel }~
\sum_{\bar m} a_{\bar m}(\varepsilon ) \Omega _{\bar m}(t) +\\
\int _{0}^{t} \frac {dt }{\parallel \Omega (t)\parallel }~
\sum _{\bar m,\bar k} \frac {\sin (\varepsilon \bar m \times\bar  
k)}{\varepsilon\bar  k^{2}}\,
\omega _{\bar m+\bar k}\,\Omega _{-\bar  m} \,\Omega _{-\bar  k}.
\end{eqnarray*}
      Applying the estimations which we had derived one arrives at
\begin{eqnarray*}
\parallel \Omega (t)\parallel \leq \left\{\parallel \Omega (0)\parallel +
tG^{1/2}(\varepsilon )\right\} +
\int _{0}^{t} K_{\sigma }{\parallel \Omega (t)\parallel }\,dt,
\end{eqnarray*}
      which by Gronwall inequality implies that on the time interval
      $(0,{\cal{T}})$ we have
\begin{eqnarray*}
\parallel \Omega (t)\parallel \leq \left\{\parallel \Omega (0)\parallel +
{\cal{T}}G^{1/2}(\varepsilon )\right\} {\rm e}^{K_{\sigma}t},
\qquad 0 \leq t \leq {\cal{T}}.
\end{eqnarray*}
     For $N \rightarrow \infty $ the difference $\Omega (0)$ between $\omega (0)$
     and ${\tilde \omega }(0)$ tends to zero, at least as $\sim N ^{-(\sigma -1)}$,
     $\parallel \Omega (0) \parallel \sim N ^{-(\sigma -1)} \rightarrow 0$,
     similarly, $G(\varepsilon ) \sim N^{- r} \rightarrow 0$, and therefore
     also $\parallel \Omega (t) \parallel \rightarrow 0$, which proves
     that the solution $\tilde \omega $ of Euler equation on $su(N)$
     approaches the corresponding hydrodynamic solution $\omega $.

     In a similar way one can prove a little more general theorem which
     summarizes our considerations.

\noindent {\bf Theorem} If
$\omega (t) \in C^{1}([0,{\cal{T}}],\,H^{\sigma }(T^2))$
is a solution of the Euler equation of hydrodynamics on the torus 
$T^2$ then corresponding
solutions ${\tilde \omega }^{N}$ (i.e. such that
${\tilde \omega }^{N}_{\bar m}(0) =$ ${\omega }^{N}_{\bar m}(0)$ for
$- \frac {N-1}{2} \leq  m_{i} < \frac {N-1}{2}$) of the $su(N)$ Euler
equations are approaching $\omega $ in the norm of $H^{k}(T^{2})$
for every $k$ satisfying $0 \leq k < \sigma - 2$.

\section{Numerical simulations}

The theorem proved in the last section shows that for large $N$ the difference between solutions of the Euler equation on $su(N)$ and the Galerkin approximation of the Euler equation  for ideal incompressible fluid is negligibly small. Numerical schemes usually applied to solve Euler equation are based on the Galerkin approximation. 
Numerical calculations were performed to test if the group-theoretic approximation could be applied to create a better numerical scheme for solving the Euler equation.
	There seems to be one important advantage of the group-theoretic finite-mode approach over the Galerkin one. Namely, the former possesses a big number of conserved quantities, which grows with $N$. The Galerkin approximation does not have this property. As a result of this difference one can expect higher stability of the group-theoretic approximation for long times.

In the computer simulations we adopted similar initial conditions as Dowker and Wolski \cite{dowker}. This means we took $\omega (\overline{x},t=0) \sim N^2\delta_{\overline{x},\overline{x}_0} -1$.
	Strictly speaking, such initial conditions are not satisfying the assumptions of our theorem. However, in both approximations, it means the Galerkin and the group- theoretic ones, these initial conditions lead to numerical solutions quite well approximated by the discrete ones. Dowker and Wolski \cite{dowker} were performing their simulations at the level of stream functions, we were doing that at the level of vorticities.

	Calculations in \cite{dowker} were done on a 
	Hewlett-Packard workstation. We agree with \cite{dowker} that it is very difficult to go to $N$s big enough for a convincing numerical illustration of the convergence proved in the theorem. The complexity of the system of ODEs strongly depends on $N$. In our case the number of equations grows linearly with $N$, the number of quadratic terms in every equation grows with $N^2$. The $N$ is limited not only by the execution time, but also, and in fact mainly, by the memory requirements. This is why we hope that our method will be much more effective in the future --- we witness every year significant progress in memory availability.

It was not our aim at this point to make calculations for the biggest possible $N$, but rather to observe the numerical stability of both methods. This is why calculations were done on a PC (Pentium II, 700MHz, 1GB RAM), in Mathematica 4.1. 
We restricted ourselves to small $N$s (relatively easy simulations). The effect is sufficiently strong to be observed at $N=11,15,25$. It is expected from theoretical considerations to be much more significant for big $N$.

		We observed individual Fourier modes as functions of time for both approximations. We were testing time after which the symmetry $\omega_{\bar k}^*=\omega_{-\bar k}$, (for our choice of initial conditions (real $\omega_{\bar 
	k}$) it was $\omega_{\bar k}=\omega_{-\bar k}$) was breaking. 
	One can easily observe big difference in times of destabilization between the two approximations. 
For $N=11$ and $N=15$ numerical calculations based on the Galerkin method destabilize after the amount of time 
approximately equal to 4.5 computer units, while the group-theoretic method  presented stable behavior 
till 20 computer time units. For $N=25$ the difference was significantly bigger. Namely, the group-theoretic method destabilized after about 18 units of time, while the Galerkin method destabilized after 1.5 units of time. In this case the group-theoretic method was more stable than the Galerkin one by a factor of 12. This difference was significant, and we are convinced that with bigger $N$ and longer simulation times the difference would increase. 
	 We used in both cases exactly the same initial conditions, numerical methods, numerical accuracy and all other parameters.

     		As an example, the results of the numerical simulations for the representative cases of $N=11,15,25$ for a pair of modes each in the Galerkin and in the group-theoretic approximations are shown in Figs.1-3. In the graphs is presented the time dependence of the amplitudes of three particular pairs of modes with the numbers $\bar k$ and $-\bar k$.
     They are representative for all the pairs of modes.
     We were looking for the instant of time when the two modes begin to behave differently showing that reliability of the simulations is lost.

\section{Conclusions}
      	In this paper we studied the 
      approximate ideal fluid description based on the
      observation concerning the convergence of the structure constants
      of the $su(N)$ ($N$ odd) algebras to the structure constants of
      the algebra {\it sdiff}$(T^2)$, and on similarity of the Euler
      equations defined on the dual algebras of $su(N)$ and {\it
      sdiff}$(T^2)$, respectively.  The hypothesis stated in the  
      literature that the Euler equations for
      $su(N)$ converge to those for {\it sdiff}$(T^2)$ was carefully examined.
      
		The results are:

	 	In addition to the fact known from the literature \cite{schl} that
	the algebras $su(\infty)$ and $sdiff(T^2)$ are  not isomorphic,
      we pointed another difficulty of the approximation. Namely, we have shown that in the limit 
      $N\to\infty$ renormalized elements of the basis  of $su(N)$, used to demonstrate
      similarity of the strucure constants of $\lim\limits_{N\to \infty} su(N)$ and of 
      $sdiff(T^2)$, become elements of infinite
      norm. 

		We proved the statement about the convergence of 
      the solutions 
      of the Euler equations for $su(N)$ to the solutions of the Euler equations for 
      $sdiff(T^2)$. This way the functional analytic part of the approximation
      eventually found a complete proof. 

		We performed extensive numerical simulations
	based on the group-theoretic and the Galerkin finite-mode approximations.
	They showed that the group-theoretic method is significantly
	more reliable for long evolution times than the Galerkin one. When the same 
	numerical conditions are applied, the Galerkin method 
	leads to instabilities after evolution time at least a few times shorter than the
	group-theoretic one. This difference grows with the number of modes used in symulation. 
For N=25  group-theoretic numerical solutions were 12 times more stable than the Galerkin ones. 
We expect the difference to be sharper for  bigger numbers of modes involved. 
On the other hand, the Galerkin method is significantly less machine 
	time and memory consuming than the group-theoretic 
      one. Galerkin method is appropriate for short evolution time calculations, performed with
 	limited computational power.  With  given machine power  at hands 
      one can obtain solutions for short evolution time using the Galerkin approximation  for 
      bigger $N$ than using the group-theoretic one.  

            The long time scale solutions of the  Euler equations are used  in meteorology, 
      in weather prediction, as well as in studying oceanic flows, to name just a few
      applications. The group-theoretic method would allow to improve long time scale predictions 
	reliability in the mentioned areas. It
	currently 
      requires powerful workstations for implementation.
      With dramatic growth of computational power and the memory size in recent years, 
	in a few years the 
      group-theoretic method should be easier to apply than it is at the
      moment.

\section{Appendix}

		Noncommutative tori are one of the simplest examples of noncommutative spaces,
	which are a subject of study in noncommutative geometry. In this appendix we
	present material on noncommutative tori, which is relevant for the finite-mode
	approximation discussed in the paper.

		The paradigm of noncommutative geometry grew up from the Gelfand-Naimark 
	theorem, accordingly to which the category of Hausdorff, locally compact 
	topological spaces is dual to 
	the category of commutative $C^*$-algebras. The commutative $C^*$-algebras
	are algebras of continuous complex-valued functions (with compact 
	support in the case of noncompact spaces) on the spaces. The $C^*$-algebras of continuous
	functions on compact spaces are additionally unital, with the unit given by the 
	constant function equal to one. Constant functions cannot have compact support
	on noncompact spaces, therefore the corresponding $C^*$-algebras are nonunital.
	Thanks to the Gelfand-Naimark theorem, the entire 
	topological information on the spaces is coded in the algebraic properties of 
	commutative $C^*$-algebras, to which they are dual.
	
		Relaxing the commutativity property on the $C^*$-algebras locates us then in the
	field of noncommutative topology. The subject of its study is not some kind of
	noncommutative spaces, but rather algebras of (continuous) functions on them.
	These "functions" are actually operators, and the $C^*$-algebras are 
 	algebras of operators acting in appropriate Hilbert spaces, in general infinite
	dimensional. This statement is the content of a second theorem by Gelfand and 
	Naimark. (On Gelfand-Naimark theorems one can consult for example the recent 
	book \cite{Varilly}). The fact that coordinates of the virtual noncommutative
	spaces are not commuting, and the possibility to realize the algebras of 
	"functions" on them as algebras of operators in a Hilbert space, inspired 
	the creators and developers of this area of mathematics to adopt physicists'
	terminology and to speak about quantum geometry or quantum topology (the latter
	name in the case of strictly limiting to the $C^*$-algebras seems more 
	appropriate) and about quantum spaces.

		In contrast to topologists, geometers usually prefer to speak about 
	smooth manifolds rather than topological spaces. Therefore, instead of $C^*$-algebras
	noncommutative geometry studies $*$-algebras that are dense in appropriate 
	$C^*$-algebras exactly in the manner in which algebras of smooth complex valued
	functions on manifolds are dense in corresponding algebras of continuous functions.
	The conditions lead to the notion of a pre-$C^*$-algebra.

		The noncommutative torus is a quantum space, which might be described using 
	either $C^*$-algebraic or $*$-algebraic language, with appropriate limits in the 
	latter case locating us in the $C^*$-algebraic framework.
		Let us shortly review, following \cite{Landi}, \cite{Michor} 
	the $*$-algebraic approach 
	to the noncommutative torus. The algebra $\cal{A}_{\theta}$ of smooth "functions"
	on the noncommutative torus $T^{2}_{\theta}$ is the unital $*$-algebra generated 
	by 2 indeterminates $U_1$, $U_2$ with the relation
\begin{equation}
		U_1 U_2 = e^{2\pi i \theta} U_2 U_1
\end{equation}
		The $*$ structure is defined by imposing unitarity on $U_1$, $U_2$, by 
	$U_1^{*} = U_{1}^{-1}$, $U_2^{*} = U_{2}^{-1}$.
	A generic element $a \in \cal{A}_{\theta}$ is then of the form
\begin{equation}
		a = \sum_{(m,n)\in \BZ^2} a_{mn} U_1^m U_2^n
\end{equation}
	, where $a_{mn}$ is a complex valued Schwarz function on $\BZ^2$, i.e. a
	sequence of complex numbers $\{ a_{mn} \in \BC : (m,n) \in \BZ^2 \}$,
	which decreases rapidly at infinity ($ \forall k \in \BN \hspace{3mm}||a||_{k} := 
	{sup}_{_{\hspace{-7mm}m,n \in \BZ}} |a_{mn}| (1 + |m| + |n|)^{k} < \infty$ ).

		In this paper we are interested in the case, in which $\theta$ is rational, 
	$\theta = M/N$, where $M$ and $N$ are positive integers taken to be relatively
	prime. Therefore we present further only the rational $\theta$ case.
	In this case ${\cal A}_{M/N}$ is the algebra of smooth sections of a 
	twisted matrix bundle over $T^2$. Let us briefly describe the construction 
	leading to the twisted bundle following \cite{Landi}. The bundle is denoted 
	${\cal M}_{q} \rightarrow T^2$, with $q = e^{2\pi i M/N}$. The fibers of the bundle
	are isomorphic with the algebra $Mat(N, \BC )$ of all $N\times N$ complex-valued matrices. 
	Its transition functions have values in $GL(N, \BC)$. They act on 
	$Mat(N, \BC)$ by conjugation. 

		First, we consider a trivial bundle (we identify here and below 
	$T^2$ with $S^1 \times S^1$)
\begin{equation}
	S^1 \times S^1 \times Mat(N, \BC) \stackrel{pr_{1,2}}{\longrightarrow} S^1 \times S^1
\label{bundle}
\end{equation}
	where $pr_{1,2}$ is the natural projection on the first two members.
	The algebra of smooth sections of this bundle is 
\begin{equation}
	C^{\infty} (S^1 \times S^1, Mat(N,\BC)) = C^{\infty}(S^1 \times S^1,\BC)\otimes Mat(N,\BC)
\end{equation}
	
		The algebra $Mat(N,\BC)$ may be understood as the unique algebra generated by 2
	unitary elements $U_0$ and $V_0$ such that 
\begin{equation}
	U_{0}V_{0} = qV_{0}U_{0}, \hspace{4mm} U_{0}^{N} = V_{0}^{N} = \B1
\end{equation}

	The matrices $U_0$ and $V_0$ may be represented in the form, which is very closely
	related to the matrices $g$ and $h$ introduced in the section 2. Here however,
	the parameter $q$ differs from the parameter $\lambda$, since $q = e^{2\pi iM/N}$,
	but $\lambda = e^{4\pi i/N}$. The algebra of smooth sections of the bundle
	(\ref{bundle}) is generated by unitary commuting elements $u$, $v$, and by unitaries 
	$U_0$, $V_0$ generating $GL(N,\BC)$.

		In order to define the twisting of this bundle, the following actions should be considered:
      the action of $\BZ_{N}\times\BZ_{N} = \BZ_{N}^2$ on $S^1\times S^1$, 
	$(u,v) \mapsto (q^m u, q^n v)$, $m,n\in \BZ_{N}$, 
	and the action of $\BZ_{N}^2$ on $Mat(N,\BC)$, $A\mapsto U_0^nV_0^{-m}AV_0^mU_0^{-n}$, $m,n\in \BZ_{N}$.

		Then, we impose the condition that the following diagram is commutative
\begin{equation}
	\begin{array}{c c c}
	S^1\times S^1\times Mat(N,\BC)
&\stackrel{\Bz_N^2}{\longrightarrow}  &{\cal M}_q\\
	pr_{1,2}\downarrow& &\downarrow p_{q}\\
	S^1\times S^1 &\stackrel{\Bz_N^2} \longrightarrow  &S^1\times S^1 
	\end{array}
\end{equation}
	
		There is a vector bundle over $S^1\times S^1$, ${\cal E}_q\rightarrow S^1 \times S^1$
	such that ${\cal M}_q = End ({\cal E}_q)$. This fact is closely related to the Morita
	equivalence of the noncommutative torus with the commutative one. Namely, Morita 
	equivalence of two $C^*$-algebras or two pre-$C^*$-algebras means that the algebras
	have a common module, on which one of them acts from the left and the other from the
	right. This is exactly the case for a noncommutative torus with rational "deformation
	parameter" and the standard commutative torus. We may treat the space of (smooth)
	sections of the bundle ${\cal E}_q\rightarrow S^1\times S^1$ as the left module for
	the algebra ${\cal M}_q$, since the latter consists of endomorphisms of the former.
	Simultaneously, since ${\cal E}_q\rightarrow S^1\times S^1$ is a bundle over the
	standard torus, the pre-$C^*$-algebra of smooth functions on the torus acts on the
	space of smooth sections of ${\cal E}_q$ from the right. This is the way in which
	the rational noncommutative tori are related via Morita equivalence with the commutative
	torus.

		However, the Morita equivalence of the two kind of spaces, although very 
	important, is not directly responsible for the convergence effects we studied in the body of
	this paper. In the geometric picture the limiting procedure may be described in the
	following way: The algebra $Mat(N, \BC)$, which is isomorphic with the fibers of the bundle
	${\cal M}_q\rightarrow S^1\times S^1$ may be undestood as the Lie algebra $gl(N, \BC)$. From this 
	moment on one should just remind the construction described in the body of the paper. 

\section{Acknowledgements}

      \paragraph{ }
	This research reported in this paper was begun and to high extend developed
	when two of the authors (HM and RO) were working at the group T-13 of
	the Theoretical Division of the Los Alamos National Laboratory, and finished
	when the two authors were working at the Environmental Science and Waste 
	Technology Division of the Laboratory. 
      The work of Z.P. was supported by Polish State Committee for
      Scientific Research (KBN) grant $N^o$ 2PO3B 169 08 1109/PO3/95/08. Z.P. 
      acknowledges    also hospitality of
      Los Alamos National Laboratory during his visit in October '97. 
	All the authors would like to express their deep gratitude to Donna 
	Spitzmiller, for her warm hospitality during their stay at Theoretical
	Division.
     
      The work of R.O. was partially supported by a fellowship from the Fulbright
      Foundation.

      H.M. was partially supported by a Directors' Funded Postdoctoral Fellowship 
	offered by Los Alamos National Laboratory.

      The authors are grateful to I.Szczyrba for very helpful
      discussions of issues concerning infinite dimensional Lie
      algebras and to K. Piech\'or for long discussions concerning
      details of the proof.
      
      The authors express their gratitude to S.Shkoller, who turned their
      attention to a recent paper \cite{Gallagher}, which was written after their
      work was completed, and which contains results on the functional analytic 
      convergence similar to ours.

\newpage

\section*{Figure captions}

\begin{description}

\item{Fig 1.1}  Group - theoretic method, N=11, the modes compared: 
\begin{description}
\item{i)} mode $(-5,0   )  $ and mode $(5,0 )   $
\item{ii)} mode $(-3,2  )   $ and mode $(3,-2   ) $
\item{iii)} mode $( -3,0 )   $ and mode $(  3,0 ) $

\end{description}
 
The simulation was still stable at 20 units of time. After this time the  $\bf k$ and $\bf -k$ modes 
started to differ from each other, which means that the solution is no longer stable at this point.

\item{Fig 1.2}  Galerkin method, N=11, the modes compared: 
\begin{description}
\item{i} mode $(-5,0  )   $ and mode $(5,0 )   $
\item{ii} mode $(-3,2  )   $ and mode $(3,-2 )   $
\item{iii} mode $ (-3,0 )   $ and mode $ ( 3,0  )$

\end{description}
The simulation was stable till 4.8  units of time. After this time the  $\bf k$ and $\bf -k$ modes 
started to differ from each other, which means that the solution is no longer stable at this point.

\item{Fig 2.1}  Group - theoretic method, N=15, the modes compared: 
\begin{description}
\item{i)} mode $(-7,0   )  $ and mode $ ( 7,0  )$
\item{ii)} mode $(-1,2 )    $ and mode $ ( 1,-2) $
\item{iii)} mode $(-7,5  )   $ and mode $ ( 7,-5) $
\end{description}
 
The simulation was stable till 23  units of time. After this time the  $\bf k$ and $\bf -k$ modes 
started to differ from each other, which means that the solution is no longer stable at this point.

\item{Fig 2.2}  Galerkin method, N=15, the modes compared: 
\begin{description}
\item{i)} mode $(-7,0 )    $ and mode $ ( 7,0 ) $
\item{ii)} mode $(-1,2  )   $ and mode $ ( 1,-2) $
\item{iii)} mode $(-7,5   )  $ and mode $(  7,-5 )$
\end{description}

The simulation was stable till 2.7  units of time.

\item{Fig 3.1}  Group-theoretic method, N=25, the modes compared: 
\begin{description}
\item{i)} mode $(-12,2    ) $ and mode $ ( 12,-2)  $
\item{ii)} mode $(-6,2    ) $ and mode $ ( 6,-2 ) $
\item{iii)} mode $(-5,2   )  $ and mode $ ( 5,-2 ) $
The simulation was stable till 18  units of time. After this time the  $\bf k$ and $\bf -k$ modes 
started to differ from each other, which means that the solution is no longer stable at this point.

\end{description}
\item{Fig 3.2}  Galerkin method, N=25, the modes compared: 
\begin{description}
\item{i)} mode $(-12,2 )    $ and mode $(  12,-2)  $
\item{ii)} mode $(-6,2 )    $ and mode $ ( 6,-2 ) $
\item{iii)} mode $(-5,2  )   $ and mode $ ( 5,-2)  $

\end{description}
The simulation was stable till 1.5  units of time. After this time the  $\bf k$ and $\bf -k$ modes 
started to differ from each other, which means that the solution is no longer stable at this point.

\end{description}

\end{document}